# Computational Design of Mixed-valence Tin Sulfides as Solar Absorbers

*Xueting Wang[1,‡], Zhun Liu[1,‡], Xin-Gang Zhao[1], Jian Lv[1], Koushik Biswas[2,*], and Lijun Zhang[1,*]*

[1]State Key Laboratory of Superhard Materials,
Key Laboratory of Automobile Materials of MOE, and School of Materials Science and Engineering, Jilin University, Changchun 130012, China

[2]Department of Chemistry and Physics, Arkansas State University, State University, AR 72467, USA

[‡]These authors contributed equally

[*]Address correspondence to: lijun_zhang@jlu.edu.cn; kbiswas@astate.edu

**Abstract**

Binary tin sulfides, such as SnS and $SnS_2$, are appealing because of their simple stoichiometry and semiconducting properties and are, therefore, being pursued as potentially cost-effective materials for optoelectronic applications. The multivalency of Sn, that is, Sn(+2) and Sn(+4) allows yet more intermediate compositions, $Sn_xS_y$, whose structures and properties are of interest. $Sn_2S_3$ is already under consideration as a mixed-valence semiconductor. Other intermediate compositions, for example, $Sn_3S_4$ and $Sn_4S_5$ have remained elusive, although their existences have been alluded to in literature. Here we report a comprehensive study of phase stability of the $Sn_xS_y$ series compounds, utilizing swarm-intelligence crystal structure search method combined with first-principles energetic calculations. We find that the stability of mixed-valence $Sn_xS_y$ compounds with respect to decomposition into pure-valence SnS and $SnS_2$ is in general weaker than the $Sn_xO_y$ counterparts, likely due to differences in chemical bonding. Besides identifying the experimentally discovered stable phases of $Sn_2S_3$, our calculations indicate that the $Sn_3S_4$ phase is another mixed-valence composition which shows marginal stability with respect to decomposition into SnS and $SnS_2$. Other studied compositions may be metastable under ambient conditions, with slightly positive formation enthalpies. We find two structures of $Sn_3S_4$ having comparably low energies, both of which feature one-dimensional chain-like fragments obtained by breaking up the edge-connected octahedral layers of $SnS_2$. Both structures indicate lattice phonon stability and one shows quasi-direct band gap with a calculated value of 1.43 eV, ideal for solar absorbers. A further analysis of the composition-structure-property relationship supports the notion that lowdimensional Sn-S motifs and van der Waals interaction may lead to diverse structure types and chemical compositions, having functional properties that are yet to be identified in the $Sn_xS_y$ series with mixed valency.

## 1. Introduction

Next generation solar absorber materials should consist of readily available and environment-friendly constituents while exhibiting excellent energy conversion efficiency.[1–4] Strong optical absorption with low Urbach energy, appropriate band edge alignment for separation of electrons and holes and high-quality pure phase materials are important considerations for solar energy conversion in photovoltaic (PV) and photocatalytic applications.[5–8] Binary tin-sulfur (Sn-S) compounds are one of the promising contenders that represent a nontoxic, earth-abundant semiconductor platform. It has a rich phase space and diverse stoichiometry, perhaps owing to the multivalency of Sn and stereochemical activity of its lone pair.[9–15]

The +2 and +4 valencies of Sn establish SnS and $SnS_2$ as prominent and relatively well-studied end members of the $Sn_xS_y$ metal-chalcogen series.[6,15,16] Under ambient conditions, SnS (*Pnma*; herzenbergite) is the known ground state, having a band gap of about 1 eV,[11,16–19] close to the optimum photon absorption threshold. At elevated temperature or pressure there are at least four known polymorphs that include orthorhombic (*Cmcm*) and cubic (zincblende, rocksalt, π-cubic) structures.[18,20–27] On the other end, $SnS_2$ is a layered two-dimensional (2D) semiconductor, which exhibits different polytypic stacking sequence depending upon growth temperature.[28,29] Its band gap, reported around 2 eV,[6,11,30–32] is large compared to SnS (*Pnma*). An intermediate compound, $Sn_2S_3$, has recently attracted attention because it features both +2 and +4 oxidation states at distinct Sn sites and possess one-dimensional (1D) chains of $SnS_2$ structural motifs with weak interchain interactions.[14,33] Electronic structure calculations predict a band gap of ~0.9 eV; however, strong absorption possibly occurs at higher values approaching 2 eV, which may explain experimental reports citing larger gap values.[33–35]

Apart from the above three compositions, indications of other stoichiometries, such as $Sn_3S_4$ and $Sn_4S_5$, are available in the literature.[28,29,36,37] Indeed, ab initio evolutionary algorithms have recently proposed new high-pressure metallic phases among tin chalcogenides, such as $Sn_3S_4$ and $Sn_3Se_4$.[38,39] The latter has been verified experimentally as described in ref 39. While metallization is not suitable for semiconductor optoelectronics, it raises the prospect of useful and potentially stable phases that remain unexplored. Intermediate $Sn_xS_y$ compounds (e.g., stable, orthorhombic $Sn_2S_3$) having multiple Sn oxidation states are associated with active lone pairs helping to stabilize distorted structures.[12,14,33] Thus, Sn-S superlattice compounds can be engineered by van der Waals (vdW) architectures with different ratio of structural motifs.[17,32,34] Recently, electron beam (E-beam) irradiation has been used for controlled removal of S atoms that mediate the transformation of few-layer $SnS_2$ to SnS.[40] Similar nonequilibrium growth may indeed stabilize new tin chalcogenide compositions, with unconventional chemistry and electronic properties. Possibility of such metastable structures – how local atomic coordination meets chemical formula and valence state requirements – remain poorly understood. We performed a variable composition structure search of the binary Sn-S system by combining swarm-intelligence algorithm and first principles calculations. The predicted structures and their energetics are consistent with previously reported ground state and metastable phases of SnS, $SnS_2$, and $Sn_2S_3$. Aside from these compositions, we report the structure and properties of new semiconducting $Sn_3S_4$ phases having low formation enthalpies and derived from 1D chain-like fragments of multivalent Sn. One potentially significant $Sn_3S_4$ phase has a predicted minimum indirect band gap of 1.43 eV, which is slightly lower than the direct gap of 1.56 eV. We discuss additional intermediate compositions that can be designed by vdW binding of 1D chain fragments and exhibiting modified electronic band structures.

## 2. Computational Details

Different $Sn_xS_y$ compositions are studied by combining crystal structure analysis with first-principles geometry optimization methods. Global optimizations to identify stable or metastable phases are carried out using the crystal structure analysis by particle swarm optimization (CALYPSO) package.[41–43] This approach requires only chemical compositions of a compound to predict stable or metastable structures at given external conditions.[41] The algorithm details and its successful applications in various material systems have been discussed elsewhere.[44–47] Additional information regarding our implementation is given in the Supporting Information. In our work, we have predicted the possible crystal structures in 13 different Sn-S stoichiometric ratios (SnS, $SnS_2$, $Sn_2S_3$, $Sn_3S_4$, $Sn_3S_5$, $Sn_4S_5$, $Sn_4S_7$, $Sn_5S_6$, $Sn_5S_7$, $Sn_5S_8$, $Sn_5S_9$, $Sn_6S_7$, $Sn_7S_8$). Structure optimization and total energy minimization are performed within density functional theory as implemented in the VASP (Vienna Ab-initio Simulation Package) code.[48] Plane-wave pseudopotential approach with generalized-gradient approximation (GGA) of Perdew, Burke, and Ernzerhof (PBE) is used.[48–50]

Since standard PBE approaches are unable to account for nonbonded dispersion forces, it is necessary to apply appropriate vdW corrections.[51–54] A recent study showed that the structural parameters of layered tin oxides described by the optB86b functional are in good agreement with the experimental results.[51,52,55–57] We adopt the optB86b vdW-DF implemented in VASP to treat dispersion interactions in the Sn-S series. The kinetic energy cutoff for the plane wave basis for all calculation was chosen to be 364 eV, with k-point mesh set to $2\pi \times 0.04$ Å$^{-1}$ for Brillouin zone integration, and a convergence criteria of -0.05 eV/Å for the force and $10^{-7}$ eV/Å for the energy. Table 1 shows the calculated lattice parameters of three ground state structures (i.e., SnS (*Pnma*), $Sn_2S_3$ (*Pnma*), and $SnS_2$ (*P-3m*1)), all of which are in good agreement – typically within 2% of reported experimental values. Lattice-dynamics calculations of the predicted phases of $Sn_3S_4$ are performed using real-space supercell approach ($2 \times 2 \times 4$ and $2 \times 4 \times 2$ optimized supercells and $3 \times 3 \times 3$ k-point mesh) as implemented in the Phonopy code.[58] Absorption coefficients are estimated via summation over occupied and unoccupied bands and evaluating the imaginary part of the dielectric tensor, while neglecting local field effects.[59]

PBE methods suffer from band gap problems due to spurious self-interaction and the lack of derivative discontinuity in the exchange-correlation potential with respect to particle occupation number.[60] As a test, we compared the band gaps obtained with modified Becke Johnson[61] (MBJ) potential to those obtained by hybrid density functional[62] (HSE06) and PBE[63] methods. The PBE gaps are significantly underestimated as expected, whereas the MBJ and HSE06 gaps of ground state structures of SnS, $Sn_2S_3$, and $SnS_2$ are in better agreement with the experimental values (Table 1). Because of the large structural and compositional search space involved in our approach, we adopt the MBJ functional as a computationally efficient route for predicting band gap values and trends. More accurate band structures of few low energy metastable phases, including those of $Sn_3S_4$, $Sn_4S_6$, $Sn_5S_8$, and $Sn_6S_{10}$, have been calculated by HSE06 functional and discussed in the following sections.

It is also known that PBE estimates of formation enthalpies of nonmetallic systems (such as oxides, sulfides, halides, etc.) are sometimes unreliable because of the incomplete cancellation of errors between the different electronic structure of the compound and elemental reference phases.[64] Several methods have been employed to correct this problem.[64-67] One such method is based on fitted elemental-phase reference energies (FERE), where the elemental reference energies are adjusted to reproduce formation enthalpies of a large number of insulators and semiconductors. It has been shown to provide accurate predictions of the thermodynamic stability and formation energies of compounds. As shown in Table 1, FERE provides better agreement with experimental formation enthalpies of the three Sn-S compositions compared to the uncorrected PBE calculations. We, thus, select FERE approach to correct the calculated formation enthalpies of predicted low-energy structures in the Sn-S system.

## 3. Results and Discussion

### 3.1. Formation Energies of the Sn-S Compounds Searched by Swarm-Intelligence.

To investigate the stability of "mixed-valence" Sn-S phases, we performed a systematic crystal structure search for $Sn_{n+m}S_{n+2m}$ compounds with the aforementioned 13 stoichiometric ratios. These potential crystals can be treated as intermediate compositions having multivalent Sn(+2) and Sn(+4) that appear between the stable end members SnS and $SnS_2$ (see Table S1). Typically, the enthalpy of formation of a binary compound $A_xB_y$ at zero temperature and pressure can be written as $\Delta H(A_xB_y) = E(A_xB_y) - xE(A) - yE(B)$, where $E(A_xB_y)$, $E(A)$, and $E(B)$ are the calculated total energy per formula unit (f.u.) of $A_xB_y$, and total energies per atom of elemental phases of A and B, respectively. When dissimilar systems are involved, such as metals (Sn) and nonmetals (S, SnS, and $SnS_2$), incomplete error calculation may lead to inaccurate $\Delta H$. For this reason, the $\Delta H$ of intermediate $Sn_{n+m}S_{n+2m}$ compounds are given in reference to the limiting phases SnS and $SnS_2$ as

$$\Delta H = [E(Sn_{n+m}S_{n+2m}) - nE(SnS) - mE(SnS_2)]/(n+m), \quad (1)$$

where $E(Sn_{n+m}S_{n+2m})$, $E(SnS)$, and $E(SnS_2)$ are the calculated total energy of $Sn_{n+m}S_{n+2m}$, SnS, and $SnS_2$, respectively. Figure 1a shows calculated $\Delta H$ of the searched $Sn_{n+m}S_{n+2m}$ compounds at 0 K, given as a function of $x = m(SnS_2)/[n(SnS)+m(SnS_2)]$.

We also applied the FERE approach (see Computational Details) to correct the formation enthalpy of the searched Sn-S compounds against their elemental phases. These are plotted in a convex hull diagram shown in Figure S1. The relative stability of phases at varying compositions are consistent with those in Figure 1a.

The ground state of the limiting phases, SnS (*Pnma*) and $SnS_2$ (*P-3m1*), are predicted successfully, which serves to verify the reliability of the structure optimization approach via swarm-intelligence. We find that the stability of mixed-valence $Sn_xS_y$ compounds with respect to decomposition into pure-valence SnS and $SnS_2$ is in general weaker than the $Sn_xO_y$ counterparts,[51] because of structural and bonding chemistry differences (discussed later). It is notable that several new SnS phases are also identified, in addition to recently proposed or experimentally observed metastable polymorphs (see Figures 1a and S2). Most of the metastable SnS crystals are built from weak interactions between distorted Sn(+2) lone pairs and neighboring S, thereby resulting in a complex local bonding and assembly geometry. The other end member, $SnS_2$, also exhibit several energetically close structures near the *P-3m*1 ground state, which are derived from layered six-coordinated Sn(+4) octahedra or prisms with interlayer slipping (see Figures 1 and S2). Our results confirm the known orthorhombic ground state of $Sn_2S_3$ (*Pnma*) [$n = 1$, $m = 1$, $x = 0.5$]. According to Figure 1a, it lies slightly above (~$3.1\times10^{-2}$ eV/f.u.) the tie-line connecting SnS and $SnS_2$, which is similar to other theory reports.[11,70] It does not necessarily denote instability w.r.t. SnS and $SnS_2$ as is evident from its successful experimental synthesis. Apart from these known $Sn_{n+m}S_{n+2m}$ compounds, we also identified low-energy structures of mixed-valent $Sn_3S_4$ [$n = 2$, $m = 1$, $x = 0.33$] with small positive ΔH around $4.3 \times 10^{-2}$ eV/f.u. (see Figure 1b-B and -B′). Few earlier studies suggested the existence of $Sn_3S_4$; however, it lacks robust experimental characterization.[28] The other studied compositions, including the $Sn_4S_5$ discussed in literature,[37,38] are not stable indicated by the higher positive $\Delta H > 67$ meV/f.u. The geometry of several lowest-energy phases and corresponding electronic properties at a few representative compositions are discussed in subsequent sections and the Supporting Information.

**3.2. Crystal Structure and Optoelectronic Properties of $Sn_3S_4$.**

Calculated lattice parameters of predicted $Sn_3S_4$ crystals are given in Table S2. The *Pbam* phase (hereafter denoted as B) adopt an orthorhombic lattice having a slightly larger volume than the monoclinic *C2/m* phase (hereafter denoted as B′). Both structures are built from 1D chain-like fragments resulting from a breakup of the edge-connected octahedral layers of $SnS_2$. As shown in Figure 1b-B and -B′, the 1D chain has an octahedrally coordinated Sn(+4) at its center. Those octahedra share edges with adjoining 1D fragments along the crystallographic c- and b-axes in B and B′ $Sn_3S_4$, respectively. In B $Sn_3S_4$, the central Sn(+4) is flanked on both sides by Sn(+2) that are coordinated with three S neighbors belonging to the same chain (Sn-S distance ~2.69-2.79 Å) and two additional S from adjacent 1D fragments (Sn-S distance ~2.96 Å) (Figure S3). The longer Sn-S bonds between neighboring chain fragments gives this crystal a three-dimensionally connected polyhedral network. In the B′ phase, the Sn(+2) end members of the 1D chain are coordinated with three S neighbors having bond lengths ~2.67-2.71 Å and adopt a somewhat symmetric trigonal pyramid arrangement. There are three additional S neighbors from adjacent chains whose distances range ~3.22-3.30 Å. The chain-like features of $Sn_3S_4$ are analogous to the longer 1D chains found in $Sn_2S_3$ which has a crossed bilayer packing to maximize the dispersion interactions. The structure of B′ $Sn_3S_4$ may be optimal for attractive dispersion interaction giving it a slightly smaller volume and may become more stable than the twisted B phase under pressure. Note that the results concerning $Sn_3S_4$ structures correspond to equilibrium volume at 0 K and separate from the high pressure, metallic $Sn_3S_4$ predicted recently by Gonzales et al.[38] Considering the high structural similarity and the small energy difference between two polymorphs, we will focus our discussion on the optoelectronic properties of B′ phase, which has a direct band gap ($E_{dir}$ = 1.56 eV) close to the Shockley-Queisser (S-Q)[2] limit for solar absorbers. The B phase, having a much lower gap ($E_{dir}$ = 0.69 eV), is briefly mentioned in the text, with additional details in the Supporting Information.

The dynamical stability of $Sn_3S_4$ crystals can be ascertained by their phonon spectra. Within the harmonic approximation the potential energy of an atom vibrating at frequency $\omega$ about the coordinate $Q = 0$ is given by $U = 1/2MQ^2\omega^2$.[58] As shown in Figure 2a, the calculated harmonic phonon spectrum at 0 K for B′ $Sn_3S_4$ exhibits no imaginary modes, that is, $\omega \geq 0$ at all wave vectors, indicating that the predicted structure is dynamically stable. Similar stability is also found in the B $Sn_3S_4$ (Figure S4a). The inclusion of vibrational free energy to the static lattice energy may further stabilize these $Sn_3S_4$ crystals against decomposition to SnS and $SnS_2$ phases at T > 0. It will be interesting to investigate the relative stability of the predicted phases by including vibrational contributions to the free energy, which is deferred to a future study.

The calculated (HSE06) electronic band structure of B′ $Sn_3S_4$ crystal is shown in Figure 2b, with an indirect band gap of about 1.43 eV. The direct gap is slightly higher, about 1.56 eV located near the D point. The gap values are close to the SQ limit which is often used as a predictor of PV cell efficiency. Perhaps more importantly, as recently highlighted by Yu and Zunger,[71] phonon-assisted absorption at the indirect gap may not be a limiting factor in this case since the direct gap is only about 0.13 eV larger than the indirect value. Because of the structural similarity between $Sn_2S_3$ and $Sn_3S_4$, both built from $SnS_2$-chain units, it is to be expected that their electronic structures have common characteristics. Similar to $Sn_2S_3$,[33] $Sn_3S_4$ has different orbital character near the valence and conduction band edges due to the mixed valence of Sn atoms. To clarify the orbital contribution near the band edges of B′ $Sn_3S_4$, the density of states along with a projection of $s$ and $p$ orbitals on the Sn(+2) and Sn(+4) atoms are shown in Figure 2c. The top of the valence bands are derived from a mixture of Sn(+2) 5$s$-S 3$p$, whereas the conduction band minimum (CBM) is dominated by Sn(+4) 5$s$ hybridized with S 3$p$ states. Thus optical excitation involves electronic transition between different Sn atoms. This spatially indirect excitation may influence the dipole transition-matrix element across the band gap.

It should be noted that the Sn(+2)-5$s$ and S-3$p$ antibonding states can rehybridize with the Sn 5$p$ states producing stereochemically active lone pairs, which further enhance the stability of the distorted structure via asymmetric electron distribution.[12] Signature of such coupling is evident from almost equal contributions of Sn(+2) 5$p$ and 5$s$ orbitals observed near the valence band maximum (VBM) in Figure 2c. A meticulous analysis of electron density contours will help to further explain the bonding interaction caused by such coupling mediated by S, possibly responsible for stabilizing this mixed-valent structure.

As shown in Figure 2d, the onset of dipole-allowed optical absorption, excluding excitonic effects, occur at photon energies near the direct electronic gap at ~1.56 eV in B′ $Sn_3S_4$, which is slightly higher than that of SnS and $Sn_2S_3$. The moderate absorption strength near the direct gap becomes stronger ($\alpha \geq 10^4$ cm$^{-1}$) when the transition is above 1.6 eV. For single-junction solar cells, it is desirable to have a small offset between the lowest indirect and the direct allowed transitions, as in the case of B′ $Sn_3S_4$. In such cases, thin-film PV application may be practicable because of strong dipole transitions in the vicinity of phonon-assisted weak absorption at the indirect gap.

We now briefly turn toward the properties of B phase $Sn_3S_4$ crystal. This structure is also found to be dynamically stable (Figure S4) and has the lowest indirect electronic band gap of 0.45 eV. The direct gap is about 0.69 eV. It indicates that the electronic states are significantly influenced by the $Sn_3S_4$ interchain sliding and twisting in comparison to the B′ phase. The reduced band gap likely originates from local bonding changes of Sn(+2). The Sn-S distances between adjacent 1D chains is shorter ~2.97 Å in B $Sn_3S_4$ compared to ~3.26 Å in B′ $Sn_3S_4$. Thus the VBM of B $Sn_3S_4$, which is composed of S $3p$ and antibonding Sn(+2) $5s$ combination, is pushed up in energy leading to higher valence bands and a smaller gap (Figure S4c). Similar to the B′ phase, the conduction edge of B $Sn_3S_4$ is derived from the Sn(+4) $5s$ and S $3p$ interactions while the states remain very dispersive. The modification of electronic states by controlling $Sn_3S_4$ chain geometry may be a new route for designing Sn-S vdW polymorphs. Such possibilities are discussed below.

**3.3. New Sn-S Compounds by Imitating the $Sn_3S_4$ Chain.**
E-beam irradiation can be an effective approach to modify 2D materials via defect engineering.[40] Several types of induced structural defects, from zero-dimensional point defects to extended domains, have been reported in layered materials, such as graphene, hexagonal boron nitride, and transition metal dichalcogenides.[11,72] In these cases the interaction of the beam with the sample can be controlled with subnanometer precision. Controlled removal of S atoms by electron irradiation to form 1D line-defects in $SnS_2$ has been successfully applied to design new Sn-S compounds.[40] It is perhaps another approach to tune band gaps and optical coefficients by introducing Sn(+2) content within $SnS_2$. Within this context, let us discuss the possibility of mixed-valence compounds assembled by varying the lengths of 1D chains and different Sn(+2) content. Figure 3a shows a mechanism for building chain-like $Sn_4S_6$ compounds by periodic removal of S atoms from the ($P$-3$m$1) $SnS_2$ supercell (removed S atoms highlighted by dashed circles around them). Notice that after removal of S atoms the optimized structure adopt a half unit glide between nearby layers, shown by the S-vacancy lines in Figure 3a. It facilitates interchain packing. Similar Sn-S homologues ($Sn_{2+m}S_{2+2m}$, for example, $Sn_3S_4$, $Sn_4S_6$, $Sn_5S_8$, and $Sn_6S_{10}$) with different chain lengths, each terminated by two Sn(+2) can be obtained from the $SnS_2$ supercells with varied line-defect concentrations. The formation energy of these Sn-S homologues is given by

$$\Delta H = [E(Sn_{2+m}S_{2+2m}) - 2E(SnS) - mE(SnS_2)]/(2+m), (2)$$

where m ≥ 0 represent the number of Sn(+4) units within a chain. The new $Sn_3S_4$ chain structure is termed as B″ $Sn_3S_4$ (separate from B and B′ $Sn_3S_4$ discussed earlier). Except for the $Sn_3S_4$ outlier, the $\Delta H$ of the Sn-S homologues shown in Figure 3b follow a gradual trend with increasing chain length. The zero of energy in Figure 3b refers to the ground states of SnS and $SnS_2$. The minimal chain-like crystal (i.e., SnS with $m = 0$ and without Sn(+4) content) is much higher in energy than its ground state herzenbergite (*Pnma*) phase. Interestingly, a similar single chain SnS polymorph was predicted by our structure search, which had a reversed interchain orientation and only 0.05 eV/f.u. above the SnS (*Pnma*) phase (see Figure S8F). In case of the longer chain Sn-S homologues (m > 2, that is, $Sn_5S_8$), the formation energies are comparable to that of the lowest energy Sn-S compounds obtained from structure search approach (see Figure S5). The deviation from the trend in Figure 3b, because of the abnormal energy reduction in B″ $Sn_3S_4$, can be understood from its interchain packing configurations. As compared in Figure S6, when decreasing the chain length from $Sn_4S_6$ to $Sn_3S_4$, the distance between two chain planes reduce from 5.64 to 4.0 Å, while the edge-to-edge distance between successive chains in the same layer increase from 3.76 to 5.38 Å. It indicates that the smaller chains allow them to slide more freely, thus enabling the structure to reach a local minimum on the energy surface. It should be noted that the formation energy of B″ $Sn_3S_4$ is still slightly higher than previously discussed B′ and B phases. The higher energy of B″ $Sn_3S_4$ possibly originates from repulsive interaction between Sn(+2) located at the chain edges which is surrounded by seven S atoms, compared to six and five S atoms in B′ and B $Sn_3S_4$, respectively.

The band structures of chain-like Sn-S compounds are similar to that of B′ $Sn_3S_4$

(thus not shown). With increasing chain length, the valence band repulsion gradually reaches a limiting value, which results in similar band gap values of the long chain Sn-S homologues. As shown in Figure 3c, the band gap of single chain SnS is about 0.79 eV, which increases to 1.44 eV in $Sn_6S_{10}$ while the gap values remain fairly uniform in the long chain compounds. Although B″ $Sn_3S_4$ have a smaller fundamental gap, its absorption spectrum is comparable to the B′ $Sn_3S_4$, as shown in Figure 3d. With increasing chain length (i.e., decreasing Sn(+2) content), the absorption onset shift toward higher energy following the trend of the increasing gap. High-performance absorber materials also require a high joint density of states (JDOS) to maximize (momentum conserving) transition near the absorption threshold. Figure 3e depicts a larger JDOS covering the energy range (~1.1-1.3 eV) in B″ $Sn_3S_4$ compared to other Sn-S homologues, including B′ $Sn_3S_4$. It implies that the spatially indirect excitation between Sn(+2)-5$s$ and Sn(+4)-5$s$ would diminish the threshold transition amplitudes in long-chain Sn-S homologues.

**3.4. Insights from the Structural and Electronic Properties of Sn-S Compounds.**

The results of the structure search emphasize the importance of basic building blocks in the Sn-S system that create an array of crystal geometries with slightly positive formation enthalpy. They are formed by vdW interaction that piece together those blocks, few of whom are highlighted in Figure S7 (marked E through J). We use two representative categories to illustrate their structural characteristics. As depicted in Figure S8, the metastable $Sn_2S_3$ (H) is constructed by different $Sn_xS_y$ compounds with 2D vdW interface (where H = E + D; see Figures 1b and S8 for structures E, D, and H). On the other hand, $Sn_4S_5$ (J) is formed by the mixed 2D and 1D vdW Sn-S chains (where J = C + 2F; see Figures 1b and S8 for structures C, F, and J). Other analogical structures designed by vdW superlattices can be engineered with the varied component ratio (see, for example, $Sn_3S_5$ (H1), and $Sn_3S_4$ (H2) in Figure S8). Although these metastable composites have positive formation enthalpy, the basic building components are energetically comparable to the ground state structures, for example, the SnS in structures E (*R-3m*) and F (*Pnma*) are only 0.016 and 0.05 eV/f.u. above SnS (*Pnma*), respectively. One may argue that the vdW binding between the Sn-S homologues (showing positive formation enthalpy) is too weak to stabilize the superstructures. The instability in the vdW heterostructures possibly arises from the cost of strain energy during structure optimization with mismatched lattice parameters, which can be relaxed by macroscopic fluctuations under experimental conditions. Albeit this computational uncertainty, it is clear that the chain/layer motif interactions in the Sn-S compounds are different from those reported in the Sn-O series.[50] In the tin oxides $Sn_xO_y$, the ground-state structures of the intermediate mixed-valence compositions are prone toward negative formation enthalpy,[50] whereas the $Sn_xS_y$ series show small positive values. Such a disparity between Sn-S and Sn-O compounds may originate from the nature of the chemical bonding between Sn-$5s$ and O/S-$p$ states. In this family of materials, interactions between the Sn-$5p$ and Sn-$5s$ orbitals occurring through the mediation of anionic O/S-$p$ states results in highly asymmetric electron density and distorted lone pairs. The strength of such interactions determines the stability of chain/layer motifs. With less electronegative S (compared to O), the orbital overlap between S-$3p$ and Sn-$5s$ is likely decreased, which indirectly results in weak Sn-$5s$ and Sn-$5p$ interaction. Thus, in contrast to the strong interaction among chain/layer motifs in the Sn-O series, the Sn-S composites adopt relatively weaker vdW binding by charge transfer interactions.

Owing to the vdW binding of basic Sn-S phases by charge transfer interactions, the interfacial dipole can significantly influence electronic band alignment. Charge redistribution in selected Sn.S vdW superlattices and their band structures are shown in Figures S9 and S10, respectively. Spatially indirect optical transitions have been observed in such vdW assembled heterostructures due to strong exciton coupling between the interfaces.[51] The built-in electric field between vdW constituents play a critical role in electronic band modulation and charge carrier extraction. However, it is not intuitive to understand the modification of electronic bands in the vdW heterostructures in relation to that of their separate building components (see Figure S10). Multiple factors may be responsible, that include coupling between the individual vdW constituents which are in turn affected by multivalent bonding and lone pairs susceptible to charge transfer.

Finally, we plot calculated band gaps and relative formation energies of the searched Sn-S crystals at different compositions (Figure 4). Though it does not reveal any direct correlation, it is notable that at least some low energy structures have similar gap values (evident from the closely distributed red and brown dots at certain compositions). This clustered behavior between formation energy and band gaps may originate from their structural similarity. For instance, most of the SnS polymorphs have triply coordinated Sn(+2) and accompanying lone pairs responsible for structural distortions. These local configurations and motifs have a nontrivial connection with corresponding electronic structures, which ultimately leads to a relationship between band gaps and formation energy. Further theoretical endeavors are expected toward establishing such a structure-property relationship among these compounds.

## 4. Summary

The binary tin sulfide compounds, namely, SnS, $SnS_2$, and $Sn_2S_3$, are deserving of attention because of their simple chemical composition and nontoxic, earth-abundant constituents. SnS is the most widely investigated material and despite its favorable electronic structure photovoltaic device efficiency has remained below 5%.[71,73] A number of loss factors may be responsible for low conversion efficiency, including intrinsic bulk properties, as well as band alignment and carrier transport at heterojunctions. While attempting to overcome these issues, an exploration of other stable compositions is also necessary to achieve enhanced performance in thin-film solar absorbers. They may be more compatible as absorber layers and amenable in terms of their bulk and surface properties. In that regard, multivalency of Sn and its lone pair activity is responsible for the rich phase space of tin sulfides, where various compositions $Sn_xS_y$ may be possible featuring Sn(+2) and Sn(+4). $Sn_2S_3$ is an example of relatively rare mixed-valence compound having desirable electronic structure for photovoltaic and thermoelectric applications. To further explore this diverse structure-composition space of $Sn_xS_y$ compounds, we have performed a comprehensive materials design study based on structure search combined with density functional calculations. It provides reliable predictions of the ground state and metastable structures of the known stable compositions of SnS, $SnS_2$, and $Sn_2S_3$. Many higher-energy metastable structures are also predicted that are yet to be experimentally verified, but may be feasible under high temperature or high-pressure growth conditions.[38,74,76] An important finding is the mixed-valent $Sn_3S_4$ composition that appears as a marginally stable phase. Its occurrence has been alluded to, but not definitively observed in previous studies. We identified structures showing dynamical stability evidenced by the absence of imaginary phonon modes. We distinguish our finding of $Sn_3S_4$ as semiconductor under ambient pressure from another recent study that predicted $Sn_3S_4$ to be metallic at a pressure of about 15 GPa or higher.[38] Although experimental verification is unavailable as of yet, our initial reports of 3:4 stoichiometry based on structure search studies suggest the presence of similar phases that may be viable. We find two low-energy structures of $Sn_3S_4$ with comparable free energies, referred in the text as B and B′ $Sn_3S_4$. The small positive formation enthalpy may be within the accuracy of our computational method. It may be possible to grow $Sn_3S_4$ phases synthesized under nonequilibrium (high pressure or temperature) conditions. We note that $Sn_2S_3$ also shows small positive enthalpy (see Figures 1 and S1), although it has been successfully synthesized. Other theory reports also found small positive enthalpy in case of $Sn_2S_3$. The B′ $Sn_3S_4$ has quasi-direct band gap with the calculated value of 1.43 eV, desirable for photovoltaic applications. The strong absorption onset is predicted at ~1.5 eV, better than that of the known mixed-valence compound of $Sn_2S_3$, which has a smaller band gap of ~1 eV but effective absorption threshold of ~1.75 eV. While these predicted properties sound promising, practical realization of these compositions and structures will require meticulous experimental efforts and perhaps additional results from theory. Our results further reveal useful design principles to leverage the structure-property relationship of this family of mixed-valence compounds. For instance, the 1D chain-like fragments that build the Sn-S homologues by removing S atoms from $SnS_2$ is also a convenient approach to tune the band gaps and optical coefficients because of the introduction of Sn(+2) fraction.

**Supporting Information**

The Supporting Information is available free of charge.

Searched $Sn_{n+m}S_{n+2m}$ compounds with 13 Sn-S stoichiometric ratios, structures information for representative compounds B, B′, B″ phase $Sn_3S_4$, the atomic modes for searched metastable SnS, $SnS_2$, (B and B′ phase) $Sn_3S_4$, and $Sn_2S_3$ structures, the formation enthalpy corrected by the FERE method for the searched Sn-S compounds against their elementary substances phases, the phonon spectra, band structure, projected density of states (PDOS) and absorption spectrum for B phase $Sn_3S_4$, the formation energies for the imitated Sn-S homologues with different chain width, the detailed atomic configurations for Sn-S homologues with increased chain width from $Sn_3S_4$ (B″) to $Sn_5S_8$ (M), the formation energy for the Sn-S compound structures built from their vdWs superstructures, the representative metastable Sn-S structure models (E-J) assembled from vdWs binding, the charge redistribution between the Sn-S vdW superlattices, and the band structures for Sn-S vdW superlattices in comparison with its separated building components.

**Acknowledgements**

We acknowledge funding support of National Natural Science Foundation of China under Grant Nos. 61722403 and 11674121, National Key Research and Development Program of China (under Grants No. 2016YFB0201204), the Recruitment Program of Global Youth Experts in China, Program for JLU Science and Technology Innovative Research Team, and the Special Fund for Talent Exploitation in Jilin Province of China. K.B. contributed to the discussion, analysis, and writing of the manuscript. Calculations were performed in part at the high performance computing center of Jilin University.

**References**

(1) Kamat, P. V. Meeting the Clean Energy Demand: Nanostructure Architectures for Solar Energy Conversion. *J. Phys. Chem. C* **2007**, *111* (7), 2834-2860.

(2) Polman, A.; Knight, M.; Garnett, E. C.; Ehrler, B.; Sinke, W. C. Photovoltaic Materials: Present Efficiencies and Future Challenges. *Science* **2016**, *352* (6283), aad4424-aad4424.

(3) Lewis, N. S. Toward Cost-Effective Solar Energy Use. *Science* **2007**, *315* (5813), 798-801.

(4) Peter, L. M. Towards Sustainable Photovoltaics: The Search for New Materials. *Philos. Trans. R. Soc. A* **2011**, *369* (1942), 1840-1856.

(5) Chen, S.; Walsh, A.; Gong, X.-G.; Wei, S.-H. Classification of Lattice Defects in the Kesterite $Cu_2ZnSnS_4$ and $Cu_2ZnSnSe_4$ Earth-Abundant Solar Cell Absorbers. *Adv. Mater.* **2013**, *25* (11), 1522-1539.

(6) Burton, L. A.; Whittles, T. J.; Hesp, D.; Linhart, W. M.; Skelton, J. M.; Hou, B.; Webster, R. F.; O'Dowd, G.; Reece, C.; Cherns, D.; et al. Electronic and Optical Properties of Single Crystal $SnS_2$: An Earth-Abundant Disulfide Photocatalyst. *J. Mater. Chem. A* **2016**, *4* (4), 1312-1318.

(7) Chen, S.; Gong, X. G.; Walsh, A.; Wei, S.-H. Electronic Structure and Stability of Quaternary Chalcogenide Semiconductors Derived from Cation Cross-Substitution of II-VI and I-III-$VI_2$ Compounds. *Phys. Rev. B: Condens. Matter Mater. Phys.* **2009**, *79* (16), 165211.

(8) Schaller, R. D.; Klimov, V. I. High Efficiency Carrier Multiplication in PbSe Nanocrystals: Implications for Solar Energy Conversion. *Phys. Rev. Lett.* **2004**, *92* (18), 186601.

(9) Burton, L. A.; Walsh, A. A Photoactive Titanate with a Stereochemically Active Sn Lone Pair: Electronic and Crystal Structure of $Sn_2TiO_4$ from Computational Chemistry. *J. Solid State Chem.* **2012**, *196*, 157-160.

(10) Ganose, A. M.; Savory, C. N.; Scanlon, D. O. Beyond Methylammonium Lead Iodide: Prospects for the Emergent Field of $ns^2$ Containing Solar Absorbers. *Chem.*

*Commun.* **2017**, *53* (1), 20-44.

(11) Kumagai, Y.; Burton, L. A.; Walsh, A.; Oba, F. Electronic Structure and Defect Physics of Tin Sulfides: SnS, Sn$_2$S$_3$ and SnS$_2$. *Phys. Rev. Appl.* **2016**, *6* (1), 014009.

(12) Walsh, A.; Watson, G. W. Influence of the Anion on Lone Pair Formation in Sn(II) Monochalcogenides: A DFT Study. *J. Phys. Chem. B* **2005**, *109* (40), 18868-18875.

(13) Skelton, J. M.; Burton, L. A.; Jackson, A. J.; Oba, F.; Parker, S. C.; Walsh, A. Lattice Dynamics of the Tin Sulphides SnS$_2$, SnS and Sn$_2$S$_3$: Vibrational Spectra and Thermal Transport. *Phys. Chem. Chem. Phys.* **2017**, *19* (19), 12452-12465.

(14) Chandrasekhar, H. R.; Mead, D. G. Long-Wavelength Phonons in Mixed-Valence Semiconductor Sn$^{II}$Sn$^{IV}$S$_3$. *Phys. Rev. B: Condens. Matter Mater. Phys.* **1979**, *19* (2), 932-937.

(15) Burton, L. A.; Walsh, A. Phase Stability of the Earth-Abundant Tin Sulfides SnS, SnS$_2$ and Sn$_2$S$_3$. *J. Phys. Chem. C* **2012**, *116* (45), 24262-24267.

(16) Vidal, J.; Lany, S.; d'Avezac, M.; Zunger, A.; Zakutayev, A.; Francis, J.; Tate, J. Band-Structure, Optical Properties, and Defect Physics of the Photovoltaic Semiconductor SnS. *Appl. Phys. Lett.* **2012**, *100* (3), 032104.

(17) Zhou, T.; Pang, W. K.; Zhang, C.; Yang, J.; Chen, Z.; Liu, H. K.; Guo, Z. Enhanced Sodium-Ion Battery Performance by Structural Phase Transition from Two-Dimensional Hexagonal-SnS$_2$ to Orthorhombic-SnS. *ACS Nano* **2014**, *8* (8), 8323-8333.

(18) Ke, F.; Yang, J.; Liu, C.; Wang, Q.; Li, Y.; Zhang, J.; Wu, L.; Zhang, X.; Han, Y.; Wu, B.; et al. High-Pressure Electrical-Transport Properties of SnS: Experimental and Theoretical Approaches. *J. Phys. Chem. C* **2013**, *117* (12), 6033-6038.

(19) Ramakrishna Reddy, K. T.; Koteswara Reddy, N.; Miles, R. W. Photovoltaic Properties of SnS Based Solar Cells. *Sol. Energy Mater. Sol. Cells* **2006**, *90* (18-19), 3041-3046.

(20) Segev, E.; Argaman, U.; Abutbul, R. E.; Golan, Y.; Makov, G. A New Cubic Prototype Structure in the IV−VI Monochalcogenide System: A DFT Study. *CrystEngComm* **2017**, *19* (13), 1751-1761.

(21) Nair, P. K.; Garcia-Angelmo, A. R.; Nair, M. T. S. Cubic and Orthorhombic SnS

Thin-Film Absorbers for Tin Sulfide Solar Cells: Cubic and Orthorhombic SnS Thin-Film Absorbers. *Phys. Status Solidi A* **2016**, *213* (1), 170-177.

(22) Parenteau, M.; Carlone, C. Influence of Temperature and Pressure on the Electronic Transitions in SnS and SnSe Semiconductors. *Phys. Rev. B: Condens. Matter Mater. Phys.* **1990**, *41* (8), 5227-5234.

(23) Chattopadhyay, T.; Pannetier, J.; Von Schnering, H. G. Neutron Diffraction Study of the Structural Phase Transition in SnS and SnSe. *J. Phys. Chem. Solids* **1986**, *47* (9), 879-885.

(24) Rabkin, A.; Samuha, S.; Abutbul, R. E.; Ezersky, V.; Meshi, L.; Golan, Y. New Nanocrystalline Materials: A Previously Unknown Simple Cubic Phase in the SnS Binary System. *Nano Lett.* **2015**, *15* (3), 2174-2179.

(25) Sun, Y.; Zhong, Z.; Shirakawa, T.; Franchini, C.; Li, D.; Li, Y.; Yunoki, S.; Chen, X.-Q. Rocksalt SnS and SnSe: Native Topological Crystalline Insulators. *Phys. Rev. B: Condens. Matter Mater. Phys.* **2013**, *88* (23), 235122.

(26) Abutbul, R. E.; Segev, E.; Zeiri, L.; Ezersky, V.; Makov, G.; Golan, Y. Synthesis and Properties of Nanocrystalline π-SnS – a New Cubic Phase of Tin Sulphide. *RSC Adv.* **2016**, *6* (7), 5848-5855.

(27) Brownson, J. R. S.; Georges, C.; Lévy-Clément, C. Synthesis of a δ-SnS Polymorph by Electrodeposition. *Chem. Mater.* **2006**, *18* (26), 6397-6402.

(28) W, A.; K, S. The P-T-X Phase Diagram of The System Sn-S. *Philips Res. Repts.* **1961**, *16*, 329-342.

(29) Pa.osz, B.; Salje, E. Lattice Parameters and Spontaneous Strain in $AX_2$ Polytypes: $CdI_2$, $PbI_2$, $SnS_2$ and $SnSe_2$. J. Appl. *Crystallogr.* **1989**, *22* (6), 622-623.

(30) Gonzalez, J. M.; Oleynik, I. I. Layer-Dependent Properties of $SnS_2$ and $SnSe_2$ Two-Dimensional Materials. *Phys. Rev. B: Condens. Matter Mater. Phys.* **2016**, *94* (12), 125443.

(31) Huang, Y.; Sutter, E.; Sadowski, J. T.; Cotlet, M.; Monti, O. L. A.; Racke, D. A.; Neupane, M. R.; Wickramaratne, D.; Lake, R. K.; Parkinson, B. A.; et al. Tin Disulfide – An Emerging Layered Metal Dichalcogenide Semiconductor: Materials Properties and Device Characteristics. *ACS Nano* **2014**, *8* (10), 10743-10755.

(32) Racke, D. A.; Neupane, M. R.; Monti, O. L. A. Valence and Conduction Band Structure of the Quasi-Two-Dimensional Semiconductor $SnS_2$. *Phys. Rev. B: Condens. Matter Mater. Phys.* **2016**, *93* (8), 085309.

(33) Singh, D. J. Optical and Electronic Properties of Semiconducting $Sn_2S_3$. *Appl. Phys. Lett.* **2016**, *109* (3), 032102.

(34) Khadraoui, M.; Benramdane, N.; Miloua, R.; Mathieu, C.; Bouzidi, A.; Sahraoui, K. Physical Properties of Sprayed $Sn_2S_3$ Nanocrystalline Thin Films. *International Journal of Applied Sciences and Innovation* **2015**, *2*, 30-37.

(35) Burton, L. A.; Colombara, D.; Abellon, R. D.; Grozema, F. C.; Peter, L. M.; Savenije, T. J.; Dennler, G.; Walsh, A. Synthesis, Characterization, and Electronic Structure of Single-Crystal SnS, $Sn_2S_3$ and $SnS_2$. *Chem. Mater.* **2013**, *25* (24), 4908-4916.

(36) Sharma, R. C.; Chang, Y. A. The S-Sn (Sulfur-Tin) System. *Bull. Alloy Phase Diagrams* **1986**, *7* (3), 269-273.

(37) Cruz, M.; Morales, J.; Espinos, J. P.; Sanz, J. XRD, XPS and Sn NMR Study of Tin Sulfides Obtained by Using Chemical Vapor Transport Methods. J. *Solid State Chem.* **2003**, *175* (2), 359-365.

(38) Gonzalez, J. M.; Nguyen-Cong, K.; Steele, B. A.; Oleynik, I. I. Novel Phases and Superconductivity of Tin Sulfide Compounds. *J. Chem. Phys.* **2018**, *148* (19), 194701.

(39) Yu, H.; Lao, W.; Wang, L.; Li, K.; Chen, Y. Pressure-Stabilized Tin Selenide Phase with an Unexpected Stoichiometry and a Predicted Superconducting State at Low Temperatures. *Phys. Rev. Lett.* **2017**, *118* (13), 137002.

(40) Sutter, E.; Huang, Y.; Komsa, H.-P.; Ghorbani-Asl, M.; Krasheninnikov, A. V.; Sutter, P. Electron-Beam Induced Transformations of Layered Tin Dichalcogenides. *Nano Lett.* **2016**, *16* (7), 4410-4416.

(41) Wang, Y.; Lv, J.; Zhu, L.; Ma, Y. CALYPSO: A Method for Crystal Structure Prediction. *Comput. Phys. Commun.* **2012**, *183* (10), 2063-2070.

(42) Wang, Y.; Lv, J.; Zhu, L.; Ma, Y. Crystal Structure Prediction via Particle-Swarm

Optimization. *Phys. Rev. B: Condens. Matter Mater. Phys.* **2010**, *82* (9), 094116.

(43) Li, Q.; Zhou, D.; Zheng, W.; Ma, Y.; Chen, C. Global Structural Optimization of Tungsten Borides. *Phys. Rev. Lett.* **2013**, *110* (13), 136403.

(44) Zhu, L.; Wang, H.; Wang, Y.; Lv, J.; Ma, Y.; Cui, Q.; Ma, Y.; Zou, G. Substitutional Alloy of Bi and Te at High Pressure. *Phys. Rev. Lett.* **2011**, *106* (14), 145501.

(45) Zhang, Y.; Wang, H.; Wang, Y.; Zhang, L.; Ma, Y. Computer-Assisted Inverse Design of Inorganic Electrides. *Phys. Rev. X* **2017**, *7* (1), 011017.

(46) Lv, J.; Wang, Y.; Zhu, L.; Ma, Y. Predicted Novel High-Pressure Phases of Lithium. *Phys. Rev. Lett.* **2011**, *106* (1), 015503.

(47) Shao, P.; Ding, L.-P.; Luo, D.-B.; Cai, J.-T.; Lu, C.; Huang, X.-F. Structural, Electronic and Elastic Properties of the Shape Memory Alloy NbRu: First-Principle Investigations. *J. Alloys Compd.* **2017**, *695*, 3024-3029.

(48) Kresse, G.; Furthmu.ller, J. Efficient Iterative Schemes for Ab Initio Total-Energy Calculations Using a Plane-Wave Basis Set. *Phys. Rev. B: Condens. Matter Mater. Phys.* **1996**, *54* (16), 11169-11186.

(49) Kresse, G.; Furthmu.ller, J. Efficiency of Ab-Initio Total Energy Calculations for Metals and Semiconductors Using a Plane-Wave Basis Set. *Comput. Mater. Sci.* **1996**, *6* (1), 15-50.

(50) Grimme, S. Semiempirical GGA-Type Density Functional Constructed with a Long-Range Dispersion Correction. *J. Comput. Chem.* **2006**, 27 (15), 1787-1799.

(51) Wang, J.; Umezawa, N.; Hosono, H. Mixed Valence Tin Oxides as Novel van Der Waals Materials: Theoretical Predictions and Potential Applications. *Adv. Energy Mater.* **2016**, *6* (1), 1501190.

(52) Fang, H.; Battaglia, C.; Carraro, C.; Nemsak, S.; Ozdol, B.; Kang, J. S.; Bechtel, H. A.; Desai, S. B.; Kronast, F.; Unal, A. A.; et al. Strong Interlayer Coupling in van Der Waals Heterostructures Built from Single-Layer Chalcogenides. *Proc. Natl. Acad. Sci. U. S. A.* **2014**, *111* (17), 6198-6202.

(53) Leung, K. K.; Wang, W.; Shu, H.; Hui, Y. Y.; Wang, S.; Fong, P. W. K.; Ding, F.; Lau, S. P.; Lam, C.; Surya, C. Theoretical and Experimental Investigations on the Growth of SnS van Der Waals Epitaxies on Graphene Buffer Layer. *Cryst. Growth Des.* **2013**, *13* (11), 4755-4759.

(54) Allen, J. P.; Scanlon, D. O.; Parker, S. C.; Watson, G. W. Tin Monoxide: Structural Prediction from First Principles Calculations with van Der Waals Corrections. J. Phys. Chem. C 2011, 115 (40), 19916-19924.

(55) Lee, K.; Murray, E.. D.; Kong, L.; Lundqvist, B. I.; Langreth, D. C. Higher-Accuracy van Der Waals Density Functional. *Phys. Rev. B: Condens. Matter Mater. Phys.* **2010**, *82* (8), 081101.

(56) Sabatini, R.; Gorni, T.; de Gironcoli, S. Nonlocal van Der Waals Density Functional Made Simple and Efficient. *Phys. Rev. B: Condens. Matter Mater. Phys.* **2013**, DOI: 10.1103/PhysRevB.87.041108.

(57) Ambrosio, F.; Miceli, G.; Pasquarello, A. Redox Levels in Aqueous Solution: Effect of van Der Waals Interactions and Hybrid Functionals. *J. Chem. Phys.* **2015**, *143* (24), 244508.

(58) Togo, A.; Oba, F.; Tanaka, I. First-Principles Calculations of the Ferroelastic Transition between Rutile-Type and $CaCl_2$-Type SiO2 at High Pressures. *Phys. Rev. B: Condens. Matter Mater. Phys.* **2008**, *78* (13), 134106.

(59) Gajdos., M.; Hummer, K.; Kresse, G.; Furthmu.ller, J.; Bechstedt, F. Linear Optical Properties in the Projector-Augmented Wave Methodology. *Phys. Rev. B: Condens. Matter Mater. Phys.* **2006**, *73* (4), 045112.

(60) Toher, C.; Filippetti, A.; Sanvito, S.; Burke, K. Self-Interaction Errors in Density-Functional Calculations of Electronic Transport. *Phys. Rev. Lett.* **2005**, *95* (14), 146402.

(61) Tran, F.; Blaha, P. Accurate Band Gaps of Semiconductors and Insulators with a Semilocal Exchange-Correlation Potential. *Phys. Rev. Lett.* **2009**, *102* (22), 226401.

(62) Krukau, A. V.; Vydrov, O. A.; Izmaylov, A. F.; Scuseria, G. E. Influence of the Exchange Screening Parameter on the Performance of Screened Hybrid Functionals. *J. Chem. Phys.* **2006**, *125* (22), 224106.

(63) Perdew, J. P.; Burke, K.; Ernzerhof, M. Generalized Gradient Approximation Made Simple. *Phys. Rev. Lett.* **1996**, *77* (18), 3865-3868.

(64) Stevanovic., V.; Lany, S.; Zhang, X.; Zunger, A. Correcting Density Functional Theory for Accurate Predictions of Compound Enthalpies of Formation: Fitted Elemental-Phase Reference Energies. *Phys. Rev. B: Condens. Matter Mater. Phys.* **2012**, *85* (11), 115104.

(65) Jain, A.; Hautier, G.; Ong, S. P.; Moore, C. J.; Fischer, C. C.; Persson, K. A.; Ceder, G. Formation Enthalpies by Mixing GGA and GGA + U Calculations. *Phys. Rev. B: Condens. Matter Mater. Phys.* **2011**, *84* (4), 045115.

(66) Wang, L.; Maxisch, T.; Ceder, G. Oxidation Energies of Transition Metal Oxides within the GGA + U Framework. *Phys. Rev. B: Condens. Matter Mater. Phys.* 2006, *73* (19), 195107.

(67) Friedrich, R.; Usanmaz, D.; Oses, C.; Supka, A.; Fornari, M.; Nardelli, M. B.; Toher, C.; Curtarolo, S. Coordination Corrected Ab Initio Formation Enthalpies. **2018**, arXiv:1811.08952. arXiv.org e-Print archive. https://arxiv.org/abs/1811.08952.

(68) Liu, H.; Chang, L. L. Y. Phase Relations in Systems of Tin Chalcogenides. *J. Alloys Compd.* **1992**, *185* (1), 183-190.

(69) Ran, F.-Y.; Xiao, Z.; Toda, Y.; Hiramatsu, H.; Hosono, H.; Kamiya, T. n-Type Conversion of SnS by Isovalent Ion Substitution: Geometrical Doping as a New Doping Route. *Sci. Rep.* **2015**, *5* (1), 10428.

(70) Price, L. S.; Parkin, I. P.; Hardy, A. M. E.; Clark, R. J. H.; Hibbert, T. G.; Molloy, K. C. Atmospheric Pressure Chemical Vapor Deposition of Tin Sulfides (SnS, $Sn_2S_3$, and $SnS_2$) on Glass. *Chem. Mater.* **1999**, *11* (7), 1792-1799.

(71) Yu, L.; Zunger, A. Identification of Potential Photovoltaic Absorbers Based on First-Principles Spectroscopic Screening of Materials. *Phys. Rev. Lett.* **2012**, *108* (6), 068701.

(72) Godinho, K. G.; Walsh, A.; Watson, G. W. Energetic and Electronic Structure Analysis of Intrinsic Defects in $SnO_2$. *J. Phys. Chem. C* **2009**, *113* (1), 439-448.

(73) Sinsermsuksakul, P.; Sun, L.; Lee, S. W.; Park, H. H.; Kim, S. B.; Yang, C.; Gordon, R. G. Overcoming Efficiency Limitations of SnS-Based Solar Cells. *Adv. Energy Mater.* **2014**, *4* (15), 1400496.

(74) Ding, L.-P.; Shao, P.; Zhang, F.-H.; Lu, C.; Ding, L.; Ning, S. Y.; Huang, X. F. Crystal Structures, Stabilities, Electronic Properties, and Hardness of MoB$_2$: First-Principles Calculations. *Inorg. Chem.* **2016**, *55* (14), 7033-7040.

(75) Ding, L.-P.; Shao, P.; Zhang, F.-H.; Lu, C.; Huang, X.-F. Prediction of Molybdenum Nitride from First-Principle Calculations: Crystal Structures, Electronic Properties, and Hardness. *J. Phys. Chem. C* **2018**, *122* (36), 21039-21046.

(76) Wang, X.; Li, Y.; Pang, Y.-X.; Sun, Y.; Zhao, X.-G.; Wang, J.-R.; Zhang, L. Rational Design of New Phases of Tin Monosulfide by First-Principles Structure Searches. *Sci. China: Phys., Mech. Astron.* **2018**, *61* (10), 107311.

**Table 1.** Experimental Lattice Parameters,[23,29,36,68] Formation Enthalpies,[15,36] and Band Gaps[15,69] in Comparison with the Computational (This Work) Results for the Selected Ground-State Structures

| Structure [a] | Lattice Parameters (Å) Exp. [this work] | | | $\Delta H$ (eV/f.u.) | | | Band Gaps (eV) | | | |
|---|---|---|---|---|---|---|---|---|---|---|
| | a | b | c | PBE | FERE | Exp. | PBE | MBJ | HSE | Exp. |
| SnS (Pnma) | 4.294 [4.27] | 4.023 [4.03] | 11.254 [11.33] | -0.94 | -1.07 | -1.04 ~ -1.02 | 0.73 | 0.91 | 1.13 | 1.08 |
| $Sn_2S_3$ (Pnma) | 8.87 [8.85] | 3.75 [3.80] | 14.02 [13.93] | -2.13 | -2.47 | -2.59 ~ -3.09 | 0.5 | 0.88 | 1.06 | 1.05 |
| $SnS_2$ (P-3m1) | 3.64 [3.68] | 3.64 [3.68] | 3.64 [3.68] | -1.22 | -1.43 | -1.54 ~ -1.89 | 1.41 | 2.23 | 2.16 | 2.25 |

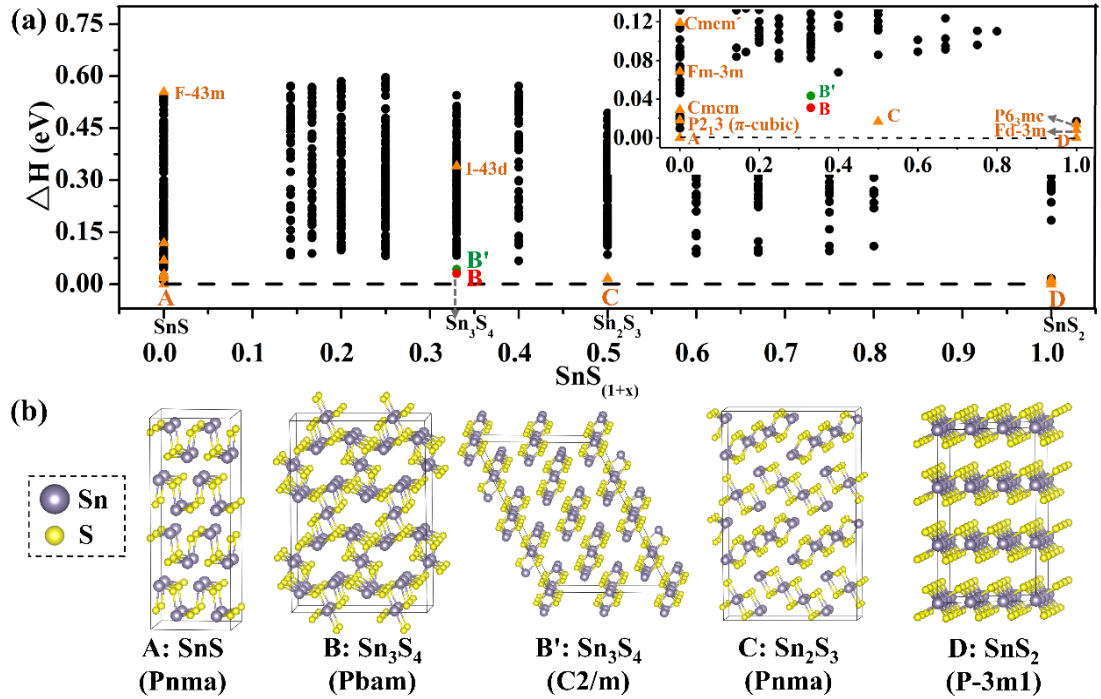

**Figure 1.** (a) Thermodynamic stability of searched compounds in the Sn-S system at 0 K plotted as a function of $x = m(SnS_2)/[n(SnS)+m(SnS_2)]$. The formation energies of $Sn_{n+m}S_{n+2m}$ compounds are referenced to the ground state of the limiting phases (SnS and $SnS_2$). Important compositions, SnS ($x = 0$), $Sn_3S_4$ ($x = 0.33$), $Sn_2S_3$ ($x = 0.5$), and $SnS_2$ ($x = 1$) are indicated along the x-axis. A magnified view of the energy axis is also shown (inset). Red and green dots represent predicted metastable structures of $Sn_3S_4$ with small positive formation energy. Orange triangles refer to the experimentally[23,25,26,31,35] or theoretically[15,20,33,39] reported structures of SnS, $SnS_2$, $Sn_2S_3$, and $Sn_3S_4$. (b) Lowest-energy crystal structures at important compositions in the Sn-S system are shown. Yellow and gray spheres denote S and Sn atoms, respectively.

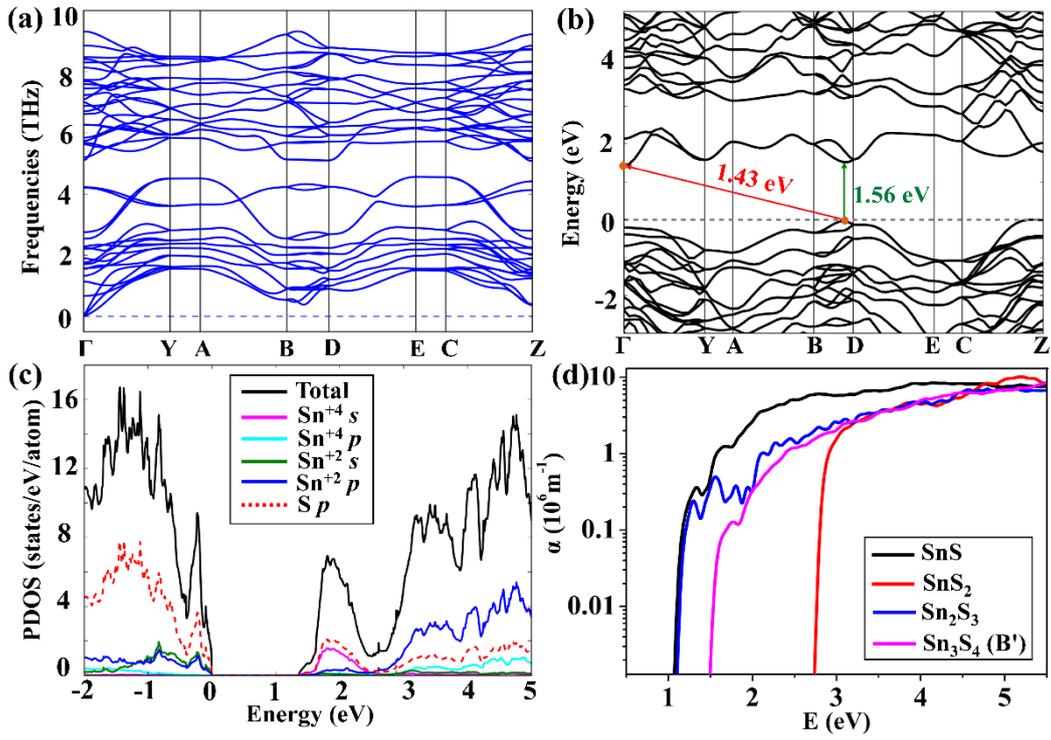

**Figure 2.** Dynamic stability, electronic, and optical properties of the B′ $Sn_3S_4$. (a) Calculated harmonic phonon spectra at 0 K. (b) Band structure and (c) projected density of states (PDOS) calculated by HSE06 functional. (d) Calculated absorption spectra of B′ $Sn_3S_4$ (magenta curve) in comparison to other Sn-S compounds

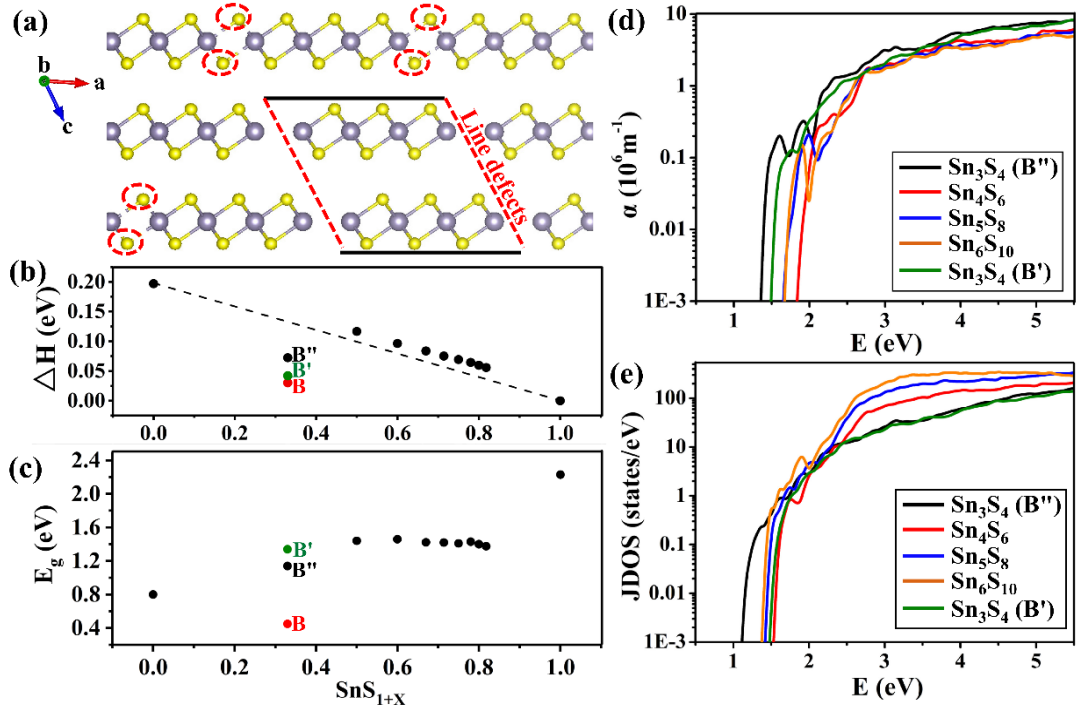

**Figure 3.** (a) Local atomic structure of a representative chain-like $Sn_4S_6$ compound created by removing S atoms from a $SnS_2$ supercell. (b) Formation enthalpies of the chain-like Sn-S homologues with different Sn(+4) fraction. Two red and green dots show the B and B′ phase of $Sn_3S_4$ obtained previously from the structure search. (c) Bandgap variations of chain-like Sn-S homologues with different Sn(+4) fraction calculated by MBJ potential. (d) Optical absorption spectra and (e) joint density of states (JDOS) of Sn-S homologues with different chain lengths and Sn(+4) fraction (the B′ $Sn_3S_4$ is also shown for comparison).

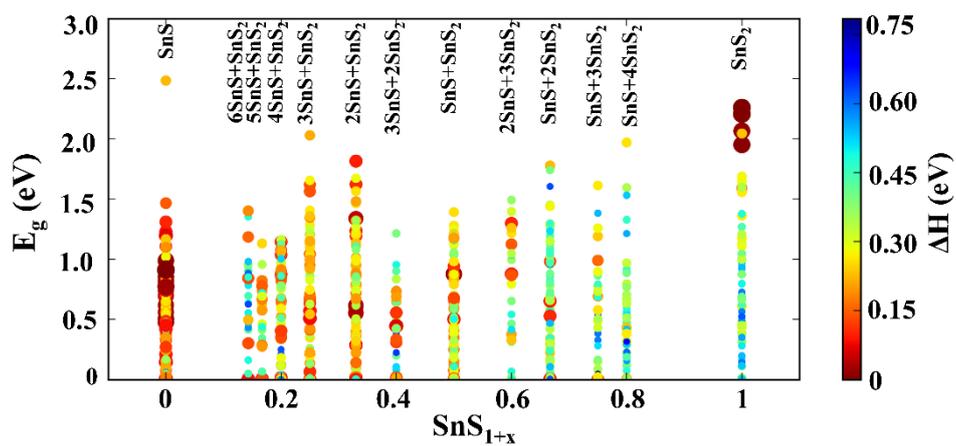

**Figure 4.** Band gaps of searched Sn-S crystals with different structures and varied stoichiometric ratios (Sn(+4) fraction) calculated by MBJ potential. The formation energies are represented by scaled colors and sizes.

**Keywords:** tin sulfides, mixed-valence semiconductor, materials by design, first-principles calculations

TOC figure

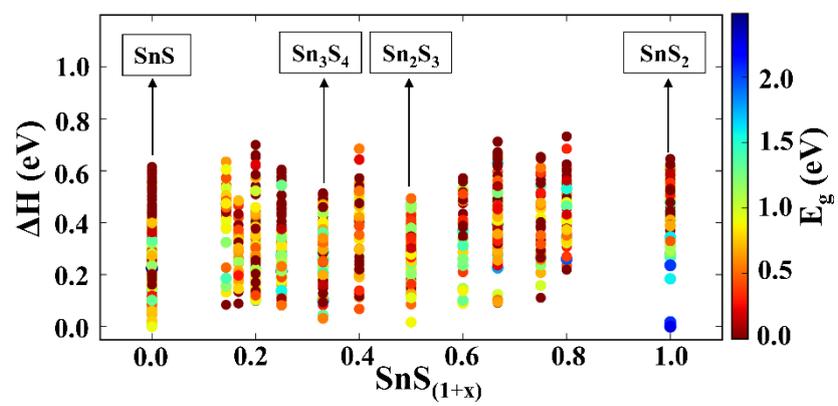

# Supporting Information

# Computational Design of Mixed-valence Tin Sulfides as Solar Absorbers

*Xueting Wang[§,‡], Zhun Liu[§,‡], Xin-Gang Zhao[§], Jian Lv[§], Koushik Biswas[†,\*], and Lijun Zhang[§,\*]*

[§]State Key Laboratory of Superhard Materials,

Key Laboratory of Automobile Materials of MOE, and College of Materials Science

and Engineering, Jilin University, Changchun 130012, China

[†]Department of Chemistry and Physics, Arkansas State University, State University,

AR 72467, USA

[‡]These authors contributed equally

[\*]Address correspondence to: lijun_zhang@jlu.edu.cn; kbiswas@astate.edu

**The details of structure searches by CALYPSO:**

The CALYPSO method comprises of four main steps. (a) Generate random structures constrained within 230 space groups. (b) Uses ab initio packages (e.g., VASP, SIESTA, and CASTEP) and force-field program (e.g., GULP) to perform the structural optimization. (c) Elimination of similar structures by using the bond characterization matrix. (d) Generate new structures by PSO for iterations.

In our work, we have performed fixed stoichiometry structure searches with system sizes containing one to four $Sn_{n+m}S_{n+2m}$ arrangements in a unit cell. For each search, the population size of each generation is 50 and it gives around 50 generations to guarantee convergence. That is, in total, 10000 ($50 \times 50 \times 4$) structures are explored for each $Sn_{n+m}S_{n+2m}$. Then we select the 200 lowest energy structures in each $Sn_{n+m}S_{n+2m}$ for more refined structural optimization. We also perform four variational stoichiometry structure prediction for further enhancing the reliability of fixed stoichiometry structure searches. For each search, the max number of atoms in the unit cell are 15, 30, 45, 60, respectively. We again select the 200 lowest energy structures in each $Sn_{n+m}S_{n+2m}$ to do refined structural optimization. Once we achieve consistency between the two procedures, similar structures are removed by the Structure Prototype Analysis Package (SPAP) and Bond Characterization Matrix (BCM).

**Table S1.** The searched crystal structures of $Sn_{n+m}S_{n+2m}$ compounds at 13 different Sn-S stoichiometric ratios. The Sn-S crystals can be treated as the intermediate compositions between SnS ($x = 0$) and $SnS_2$ ($x = 1$) where, $x = m(SnS_2)/[n(SnS)+m(SnS_2)]$.

| Sn-S compounds | SnS | $Sn_7S_8$ | $Sn_6S_7$ | $Sn_5S_6$ | $Sn_4S_5$ | $Sn_3S_4$ | $Sn_5S_7$ |
|---|---|---|---|---|---|---|---|
| $mSnS_2/[nSnS+mSnS_2]$ | 0 | 0.143 | 0.167 | 0.2 | 0.25 | 0.33 | 0.4 |
| Sn-S compounds | $Sn_2S_3$ | $Sn_5S_8$ | $Sn_3S_5$ | $Sn_4S_7$ | $Sn_5S_9$ | $SnS_2$ | |
| $mSnS_2/[nSnS+mSnS_2]$ | 0.5 | 0.6 | 0.667 | 0.75 | 0.8 | 1 | |

**Table S2.** Structural data of the B, B', B" $Sn_3S_4$ phases.

| Compounds Space group | Volume per atom (Å³/atom) | Lattice parameters (Å) | Wyckoff positions | Atoms | x | y | z |
|---|---|---|---|---|---|---|---|
| B: $Sn_3S_4$ (Pbam) | 25.55 | a=11.39 | 1a | Sn1 | 0.000 | 0.500 | 0.500 |
| | | b=7.93 | 1a | Sn2 | 0.500 | 0.000 | 0.500 |
| | | c=3.88 | 1a | Sn3 | 0.205 | 0.083 | 0.000 |
| | | | 1a | Sn4 | 0.795 | 0.917 | 0.000 |
| | | | 1a | Sn5 | 0.295 | 0.583 | 0.000 |
| | | | 1a | Sn6 | 0.705 | 0.417 | 0.000 |
| | | | 1a | S1 | 0.428 | 0.199 | 0.000 |
| | | | 1a | S2 | 0.572 | 0.801 | 0.000 |
| | | | 1a | S3 | 0.072 | 0.699 | 0.000 |
| | | | 1a | S4 | 0.928 | 0.301 | 0.000 |
| | | | 1a | S5 | 0.297 | 0.866 | 0.500 |
| | | | 1a | S6 | 0.703 | 0.134 | 0.500 |
| | | | 1a | S7 | 0.203 | 0.366 | 0.500 |
| | | | 1a | S8 | 0.797 | 0.634 | 0.500 |
| B': Sn₃S₄ (C2/m) | 25.02 | a=12.30 b=3.81 c=8.74 β=117° | 1a | Sn1 | 0.000 | 0.000 | 0.000 |
| | | | 1a | Sn2 | 0.794 | 0.000 | 0.288 |
| | | | 1a | Sn3 | 0.206 | 0.000 | 0.712 |
| | | | 1a | Sn4 | 0.500 | 0.500 | 0.000 |
| | | | 1a | Sn5 | 0.294 | 0.500 | 0.288 |
| | | | 1a | Sn6 | 0.706 | 0.500 | 0.712 |
| | | | 1a | S1 | 0.128 | 0.500 | 0.955 |
| | | | 1a | S2 | 0.872 | 0.500 | 0.045 |
| | | | 1a | S3 | 0.849 | 0.000 | 0.685 |
| | | | 1a | S4 | 0.151 | 0.000 | 0.315 |
| | | | 1a | S5 | 0.628 | 0.000 | 0.955 |
| | | | 1a | S6 | 0.372 | 0.000 | 0.045 |
| | | | 1a | S7 | 0.349 | 0.500 | 0.685 |
| | | | 1a | S8 | 0.651 | 0.500 | 0.315 |
| B": Sn₃S₄ (C2/m) | 26.19 | a=12.18 b=3.82 c=12.14 β=141° | 1a | Sn1 | 0.000 | 0.000 | 0.000 |
| | | | 1a | Sn2 | 0.500 | 0.500 | 0.000 |
| | | | 1a | Sn3 | 0.571 | 0.500 | 0.713 |
| | | | 1a | Sn4 | 0.429 | 0.500 | 0.287 |
| | | | 1a | Sn5 | 0.071 | 0.000 | 0.713 |
| | | | 1a | Sn6 | -0.071 | 0.000 | 0.287 |
| | | | 1a | S1 | 0.107 | 0.500 | -0.048 |
| | | | 1a | S2 | 0.893 | 0.500 | 0.048 |
| | | | 1a | S3 | 0.607 | 0.000 | -0.048 |
| | | | 1a | S4 | 0.393 | 0.000 | 0.048 |
| | | | 1a | S5 | 0.702 | 0.000 | 0.681 |
| | | | 1a | S6 | 0.298 | 0.000 | 0.319 |
| | | | 1a | S7 | 0.202 | 0.500 | 0.681 |
| | | | 1a | S8 | 0.798 | 0.500 | 0.319 |

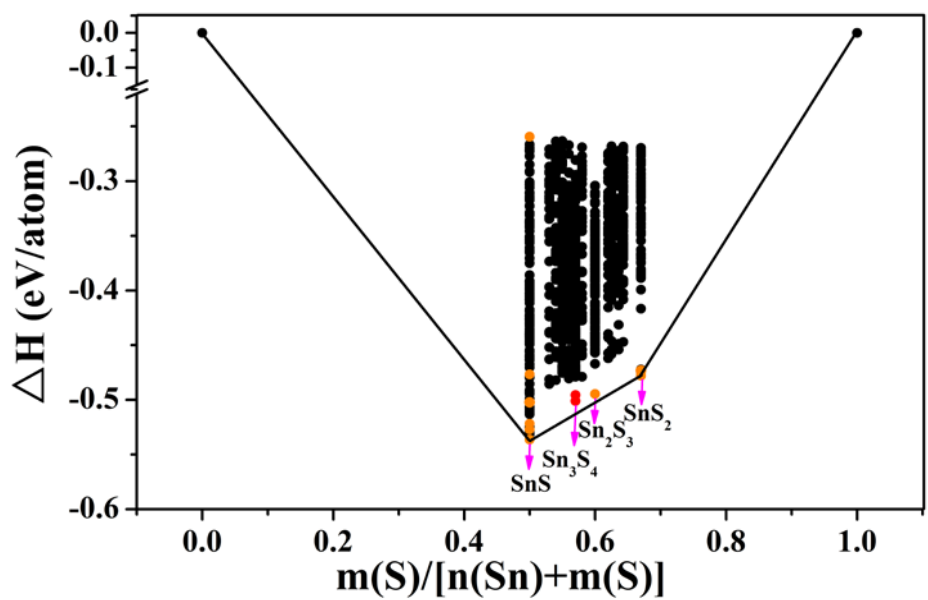

**Figure S1.** Convex hull showing the phase stability of several compounds resulting from the search of the Sn-S system. The formation enthalpies are calculated by FERE method.

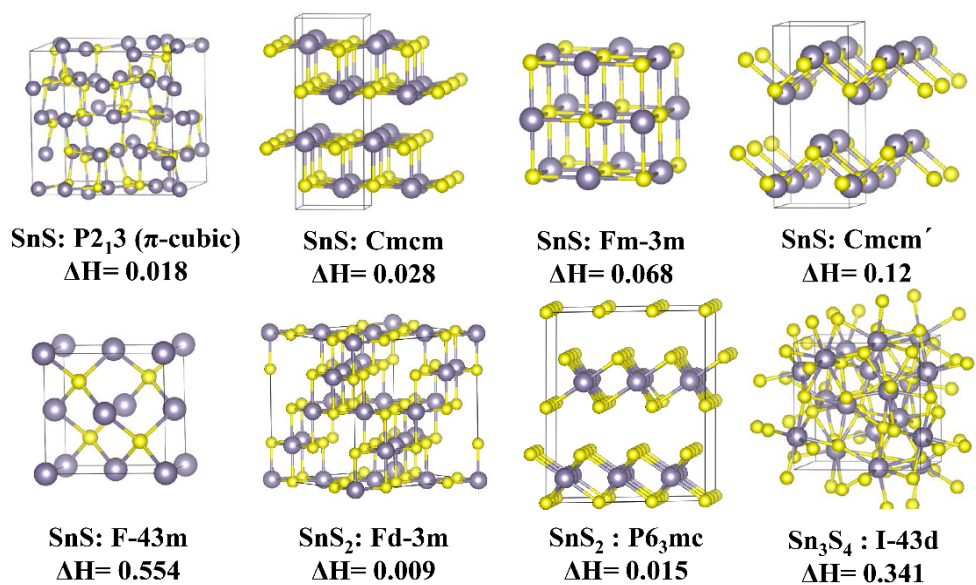

| SnS: P2$_1$3 (π-cubic) | SnS: Cmcm | SnS: Fm-3m | SnS: Cmcm´ |
| ΔH= 0.018 | ΔH= 0.028 | ΔH= 0.068 | ΔH= 0.12 |

| SnS: F-43m | SnS$_2$: Fd-3m | SnS$_2$ : P6$_3$mc | Sn$_3$S$_4$ : I-43d |
| ΔH= 0.554 | ΔH= 0.009 | ΔH= 0.015 | ΔH= 0.341 |

**Figure S2.** The searched metastable SnS and SnS$_2$ structures that have been reported in previous experiments.[1–9]

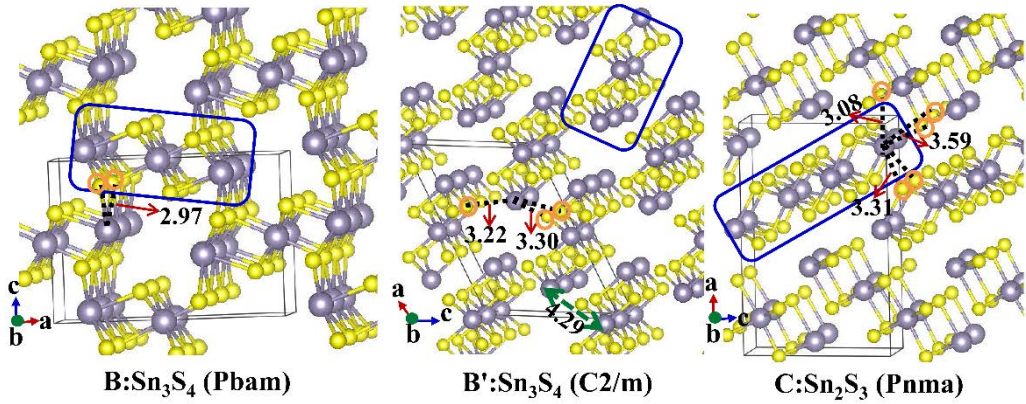

**Figure S3.** The detailed atomic bonding configurations of the B and B' phase $Sn_3S_4$ and C $Sn_2S_3$. Along the (010) plane, the Sn atoms bridged with two S atoms from nearby chains are indicated by dashed lines.

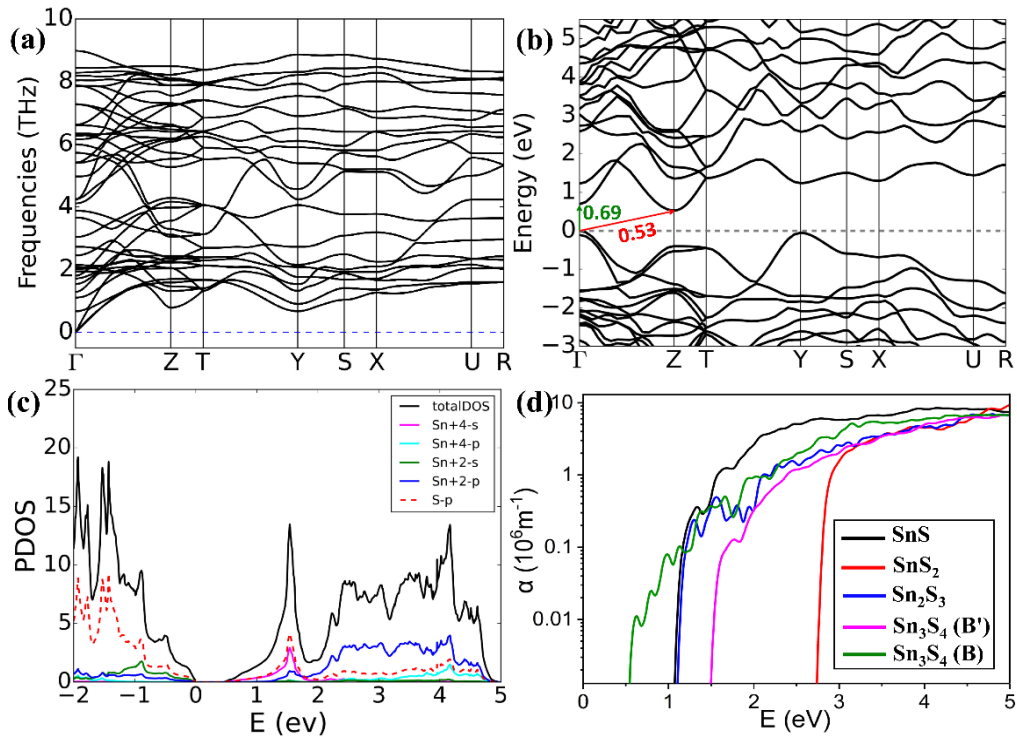

**Figure S4.** Calculated harmonic phonon spectra at 0K of B $Sn_3S_4$ (a). Corresponding band structure (b), and the projected density of states (PDOS) (c) are calculated by HSE06 potential, (d) Calculated absorption spectra of B $Sn_3S_4$ (green curve) shown in comparison to other Sn-S compounds.

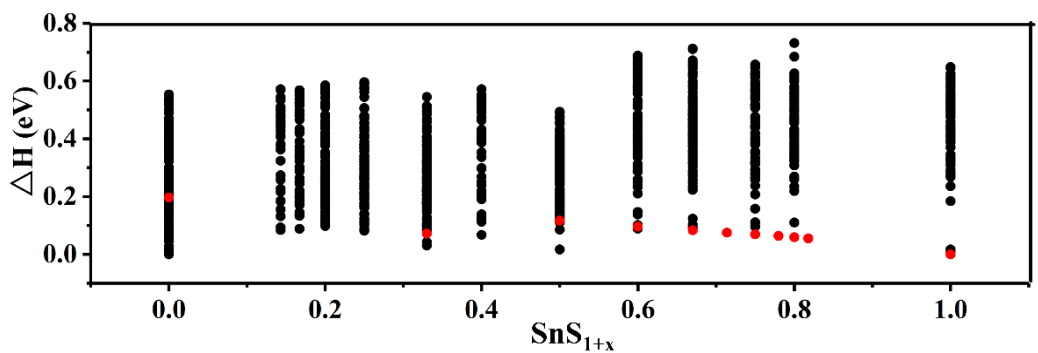

**Figure S5.** Formation energies of the intermediate Sn-S homologs (Section 3.3 in main text) with different chain length (shown as red dots) are comparable to the lowest energy structures obtained from our searched Sn-S compounds.

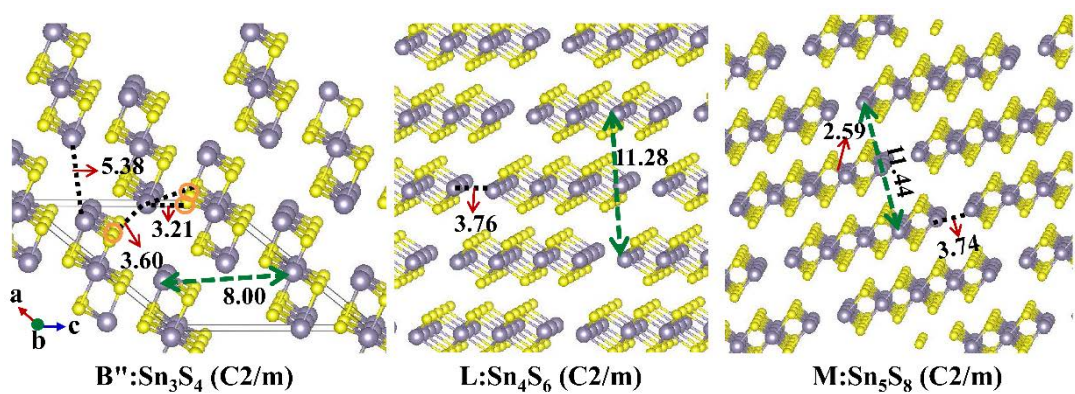

**Figure S6.** Detailed atomic configurations of Sn-S homologs with increasing chain length from $Sn_3S_4$ (B") to $Sn_5S_8$ (M).

**Sn-S compounds assembled by van der Waals binding**

Most of the intermediate stoichiometric ratios of Sn-S compounds are formed by vdW binding of the basic Sn-S phases. Form our searched results, two representative categories are used for illustration of the structure characteristic. As depicted in Figure S8, one is constructed by the typical 2D vdW interface (E+D), e.g. the searched metastable $Sn_2S_3$ (H) and $Sn_4S_5$ (J); and the other is formed by the mixed 2D and 1D vdW Sn-S chains (C+2F). The Sn-S allotropes designed from vdW superlattices can be also engineered by the analogical structures with the varied component ratio, see the $Sn_3S_5$ (H1), and $Sn_3S_4$ (H2).

We further clearly show the charge redistribution between the Sn-S vdW superlattices in Figure S9. The electrons depletion from the Sn(+2) lone pairs and accumulation to the nearby S atoms leads to an interfacial dipole. Such charge transfer induced dipole can significantly influence the band alignment in the low dimensional materials due to the quantum capacitance. Spatially indirect optical transitions have observed in such vdW assembled heterostructures due to the strong exciton coupling between the interfaces. The built-in electric field between the vdW constituents plays a critical role in electronic band modulation and charge carrier extraction. As compared in Figure S10, the band structures of the Sn-S vdW superlattices have significantly affected by the band offsets between the separated vdW building components. Take $Sn_2S_3$ (H) as an example, the local charge transfer from Sn(+2) to the $SnS_2$ give rise to a fluctuated potential step between the interfaces, which splits the degenerated $SnS_2$ bands and downward shifts the $SnS_2$ energy levels to form a type II band alignment. Apart from this contact potential effects, the accumulated electrons on the S atoms can also induce band renormalization from the reduced the Sn(+4)-S hybridization, as indicated by the less dispersive conduction bands in the heterostructures. Thus it is not intuitive to understand the electronic bands modified by the multiple factors in the Sn-S series with strong coupling between the vdW constituents, which are not only affected by its multivalent bonding but also by the susceptible lone pairs with variable spatial charge transfer.

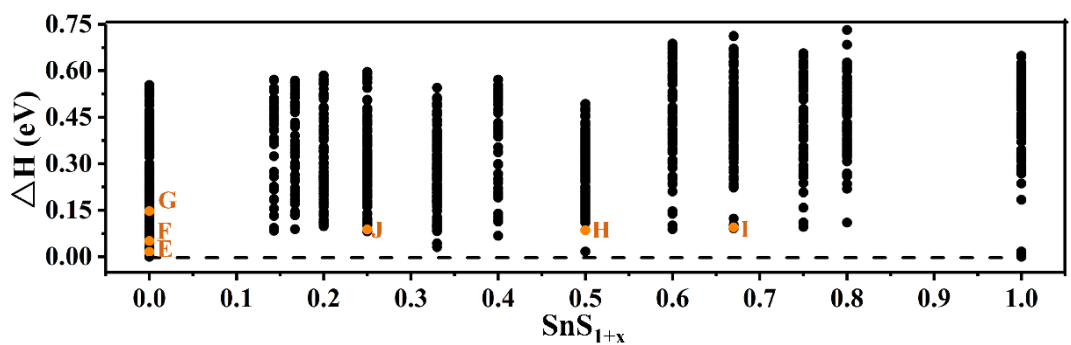

**Figure S7.** The orange (E through J) dots indicate the formation energies of the basic Sn-S compound structures and their vdW superstructures.

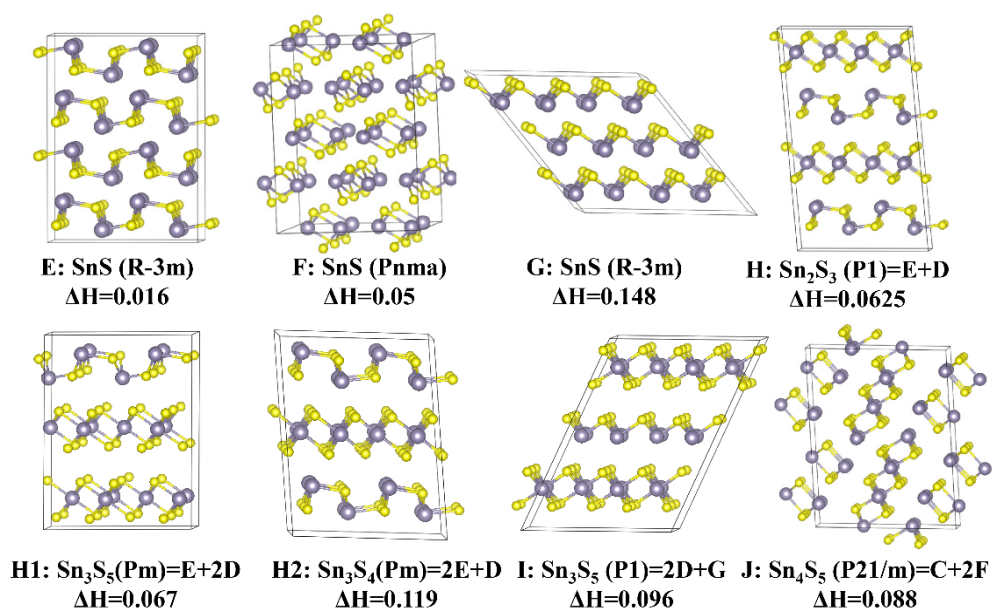

**Figure S8.** Searched representative metastable Sn-S structures (E-J) assembled through vdW binding. H1 and H2 are analogous to H with the varied component ratio.

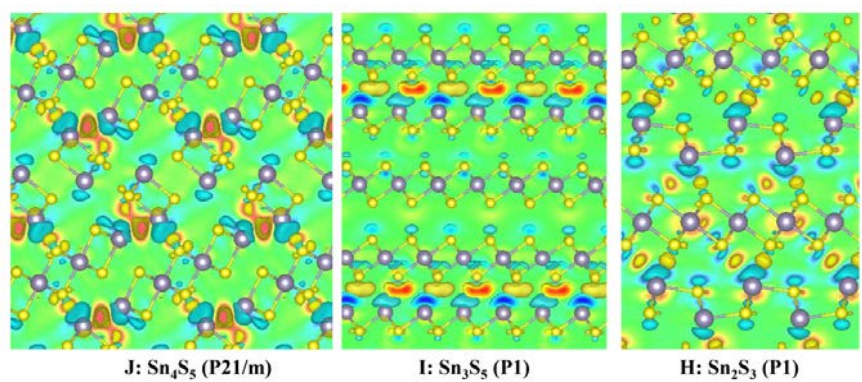

**Figure S9.** Charge redistribution between the Sn-S vdW superlattices, taking J, I and H crystals as examples.

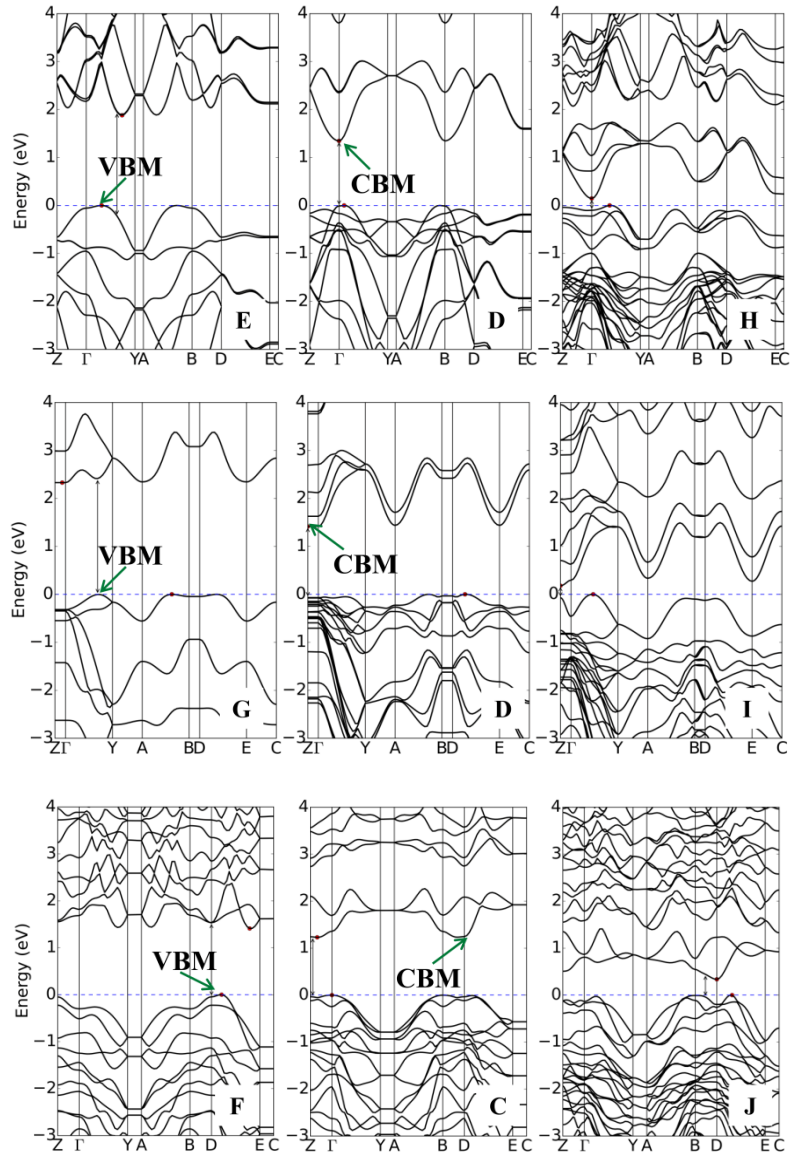

**Figure S10.** The band structures of Sn-S vdW superlattices marked by H, I, J (lower right corner) are compared with their separate building components marked by C, D, E, F G. The crystal geometry of all structures are shown in this document and main text.

## References

(1) Gonzalez, J. M.; Nguyen-Cong, K.; Steele, B. A.; Oleynik, I. I. Novel Phases and Superconductivity of Tin Sulfide Compounds. *The Journal of Chemical Physics* **2018**, *148* (19), 194701.

(2) Vidal, J.; Lany, S.; d'Avezac, M.; Zunger, A.; Zakutayev, A.; Francis, J.; Tate, J. Band-Structure, Optical Properties, and Defect Physics of the Photovoltaic Semiconductor SnS. *Applied Physics Letters* **2012**, *100* (3), 032104.

(3) Segev, E.; Argaman, U.; Abutbul, R. E.; Golan, Y.; Makov, G. A New Cubic Prototype Structure in the IV-VI Monochalcogenide System: A DFT Study. *CrystEngComm* **2017**, *19* (13), 1751-1761.

(4) Kumagai, Y.; Burton, L. A.; Walsh, A.; Oba, F. Electronic Structure and Defect Physics of Tin Sulfides: SnS, $Sn_2S_3$, and $SnS_2$. *Physical Review Applied* **2016**, *6* (1), 014009.

(5) Singh, D. J. Optical and Electronic Properties of Semiconducting $Sn_2S_3$. *Applied Physics Letters* **2016**, *109* (3), 032102.

(6) Burton, L. A.; Walsh, A. Phase Stability of the Earth-Abundant Tin Sulfides SnS, $SnS_2$, and $Sn_2S_3$. *The Journal of Physical Chemistry C* **2012**, *116* (45), 24262-24267.

(7) Burton, L. A.; Colombara, D.; Abellon, R. D.; Grozema, F. C.; Peter, L. M.; Savenije, T. J.; Dennler, G.; Walsh, A. Synthesis, Characterization, and Electronic Structure of Single-Crystal SnS, $Sn_2S_3$, and $SnS_2$. *Chemistry of Materials* **2013**, *25* (24), 4908–4916.

(8) Huang, Y.; Sutter, E.; Sadowski, J. T.; Cotlet, M.; Monti, O. L. A.; Racke, D. A.; Neupane, M. R.; Wickramaratne, D.; Lake, R. K.; Parkinson, B. A.; et al. Tin Disulfide-An Emerging Layered Metal Dichalcogenide Semiconductor: Materials Properties and Device Characteristics. *ACS Nano* **2014**, *8* (10), 10743-10755.

(9) Liu, X.; Zhao, H.; Kulka, A.; Trenczek-Zając, A.; Xie, J.; Chen, N.; Świerczek, K. Characterization of the Physicochemical Properties of Novel $SnS_2$ with Cubic Structure and Diamond-like Sn Sublattice. *Acta Materialia* **2015**, *82*, 212-223.